\documentclass[journal]{IEEEtran}
\usepackage{cite}
\usepackage{graphicx, epstopdf}
\usepackage{amsmath, amsfonts}
\usepackage{multirow, threeparttable}
\usepackage{textcomp}

\begin{document}
\title{Experimental Results of a 3D Millimeter-Wave Compressive-Reflector-Antenna Imaging System}

\author{Weite Zhang, Ali Molaei, Juan Heredia-Juesas, Luis Tirado, Katherine Graham, A. Bisulco, \\ Hipolito Gomez-Sousa, and Jose A. Martinez-Lorenzo
\thanks{W. Zhang, A. Molaei, L. Tirado, and A. Bisulco are with the Department of Electrical and Computer Engineering, Northeastern University, Boston, MA, 02115 USA.}
\thanks{K. Graham is with the Department of Mechanical and Industrial Engineering, Northeastern University, Boston, MA, 02115 USA.}
\thanks{Juan Heredia-Juesas and J. A. Martinez-Lorenzo are with the Department of Electrical and Computer Engineering and Department of Mechanical and Industrial Engineering, Northeastern University, Boston, MA, 02115 USA (e-mail: j.martinez-lorenzo@northeastern.edu).}
}

\markboth{}
{}
\maketitle

\begin{abstract}
This letter presents the first experimental results of our three-dimensional (3D) millimeter-wave (mm-wave) Compressive-Reflector-Antenna (CRA) imaging system. In this prototype, the CRA is 3D-printed and coated with a metallic spray to easily introduce pseudo-random scatterers on the surface of a traditional reflector antenna (TRA). The CRA performs a pseudo-random coding of the incident wavefront, thus adding spatial diversity in the imaging region and enabling the effective use of compressive sensing (CS) and imaging techniques. The CRA is fed with a multiple-input-multiple-output (MIMO) radar, which consists of four transmitting and four receiving ports. Consequently, the mechanical scanning parts and phase shifters, which are necessary in conventional physical or synthetic aperture arrays, are not needed in this system. Experimental results show the effectiveness of the prototype to perform a successful 3D reconstruction of a T-shaped metallic target.
\end{abstract}


\IEEEpeerreviewmaketitle

\section{Introduction}
\begin{figure}[t]
\centering
\includegraphics[width=\linewidth]{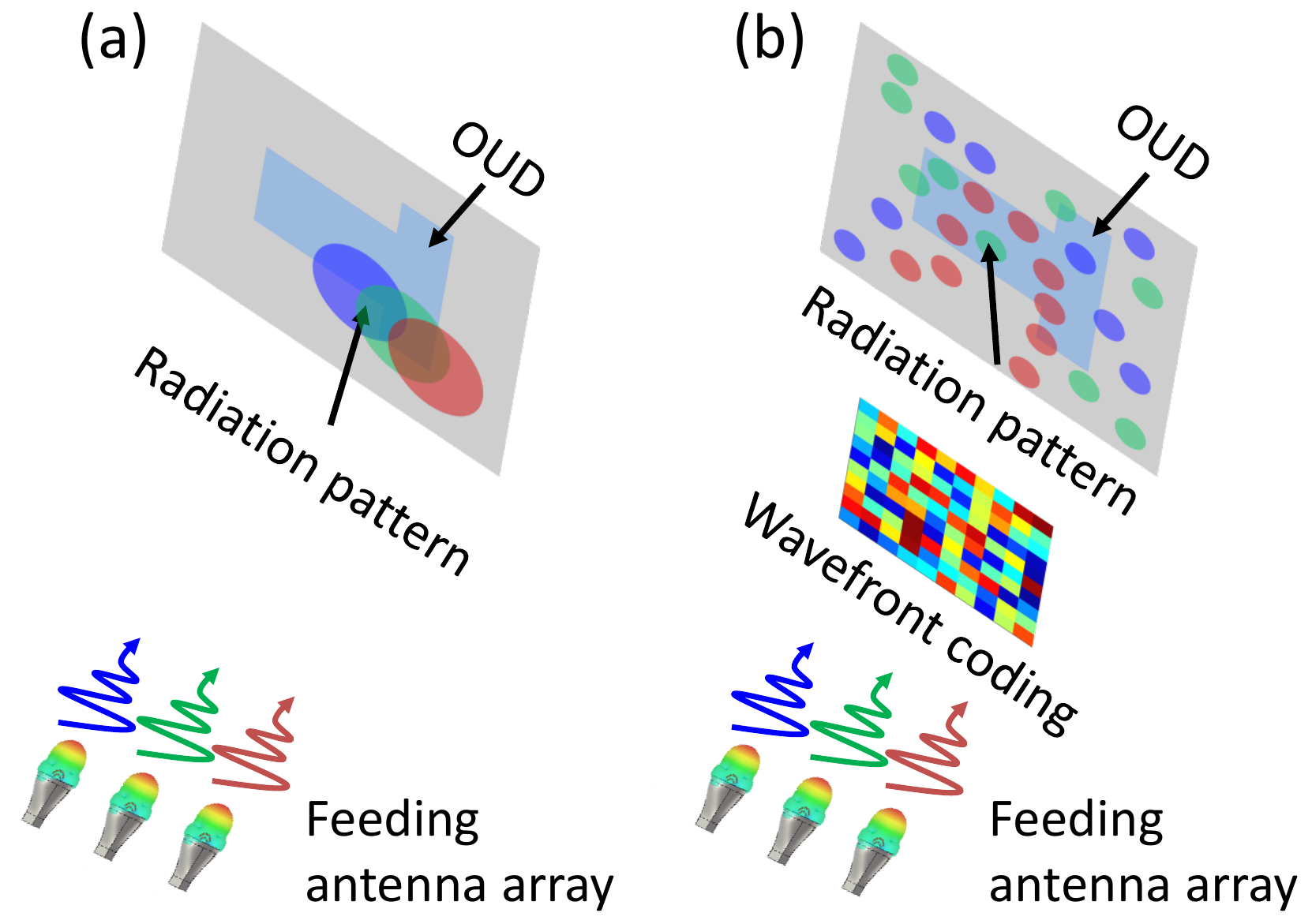}
\caption{(a) Conventional imaging system and (b) CS imaging system. OUT: object under detection, which is T-shaped.}
\label{concept_spatial_coding}
\end{figure}

\IEEEPARstart{A}{ctive} micro-wave and millimeter-wave (mm-wave) systems have been widely used in many applications, such as communications,  non-destructive testing, security screening, medical diagnosis, and through-the-wall imaging \cite{5552608, Ahmed2012Advanced, Martinez2012SAR, Meaney2000clinical, 6387638, martinez2006shapedS}. These applications build upon the unique features provided by electromagnetic waves at these frequencies; including but not limited to their ability to penetrate through optically opaque materials and their safe use around persons afforded by the non-ionizing radiation. Unfortunately, conventional high-resolution mm-wave imaging systems require the use of either expensive physical arrays\cite{lopez20003, Sheen2001Three, gonzalez2016improvingS} or slow synthetic\cite{patel2010compressed, yang2013random} aperture arrays, thus preventing its pervasive usage in societally important applications. Moreover, conventional phased array imaging systems--see Fig.\ref{concept_spatial_coding}(a)--are suboptimal due to the often large mutual information existing between successive measurements\cite{7857052}. One way to address the aforementioned limitations is to engineer new mm-wave imaging systems capable of making that successive measurements are as incoherent as possible; this will enable real-time operation while reducing the number of measurements, hardware complexity, and required energy budget.

Compressive sensing (CS) theory\cite{donoho2006compressed, baraniuk2007compressive} has been widely used to reconstruct signals from undersampled data. CS requires both the unknown signal to be sparse in a given dictionary or basis functions and the sensing matrix--relating the unknown signal and undersampled data--to satisfy the so-called Restricted Isometric Property (RIP). In many radar applications, induced sources on the surface of realistic targets can often be approximated by sparse signals\cite{8260873}, thus making a CS framework well suited to be used for imaging in such applications. Under these considerations, the target would be efficiently reconstructed with a small number of measurements.

Introducing artificial structures between the multiple-input-multiple-output (MIMO) array and the imaging region can provide an enhanced spatial diversity, as shown in Fig. \ref{concept_spatial_coding}(b); and, as a result of this coding, the number of measurements needed to obtain an acceptable image reconstruction can be reduced even more \cite{8064182}.  Many of these pioneering compressive imaging systems used spatially modulating devices--like dielectric lenses\cite{popoff2010measuring, liutkus2014imaging} or metasurfaces\cite{liu2016flexible, gollub2017large}--to perform such a wave-field-based analog data compression.

This paper presents the first experimental results of our compressive imaging system, which performs the coding using a compressive reflector antenna (CRA) \cite{7289386,7928846}. The CRA is manufactured by coating the surface of a traditional reflector antenna (TRA) with  pseudo-random metallic applique scatterers\cite{7289386}. As a result, the incident wave from the transmitting (Tx) array is spatially coded and, later, reflected towards the Region of Interest (RoI). The pseudo-random spatial code created by the CRA in the RoI is dynamically changed by selecting different input frequencies and  active ports in the MIMO array. This CRA-based system reduces the mutual information between successive measurements when compared to that of a TRA. The latter results in an enhanced sensing capacity--a metric used to quantitatively determine the information transfer efficiency between sensors and RoI \cite{7928846}-- thus potentially enabling real-time imaging while reducing the hardware cost and complexity. This letter is organized as follows: in Section II, the design details of the CRA are described; in Section III, a 3D mm-wave experimental setup is used to image a T-shaped metallic target, and the 3D reconstructed images show the reliability of the established cost-effective imaging system; and a conclusion is finally drawn in Section IV.

\section{Design and fabrication of the CRA-based Imaging System}
\begin{figure}[t]
\centering
\includegraphics[width=\linewidth]{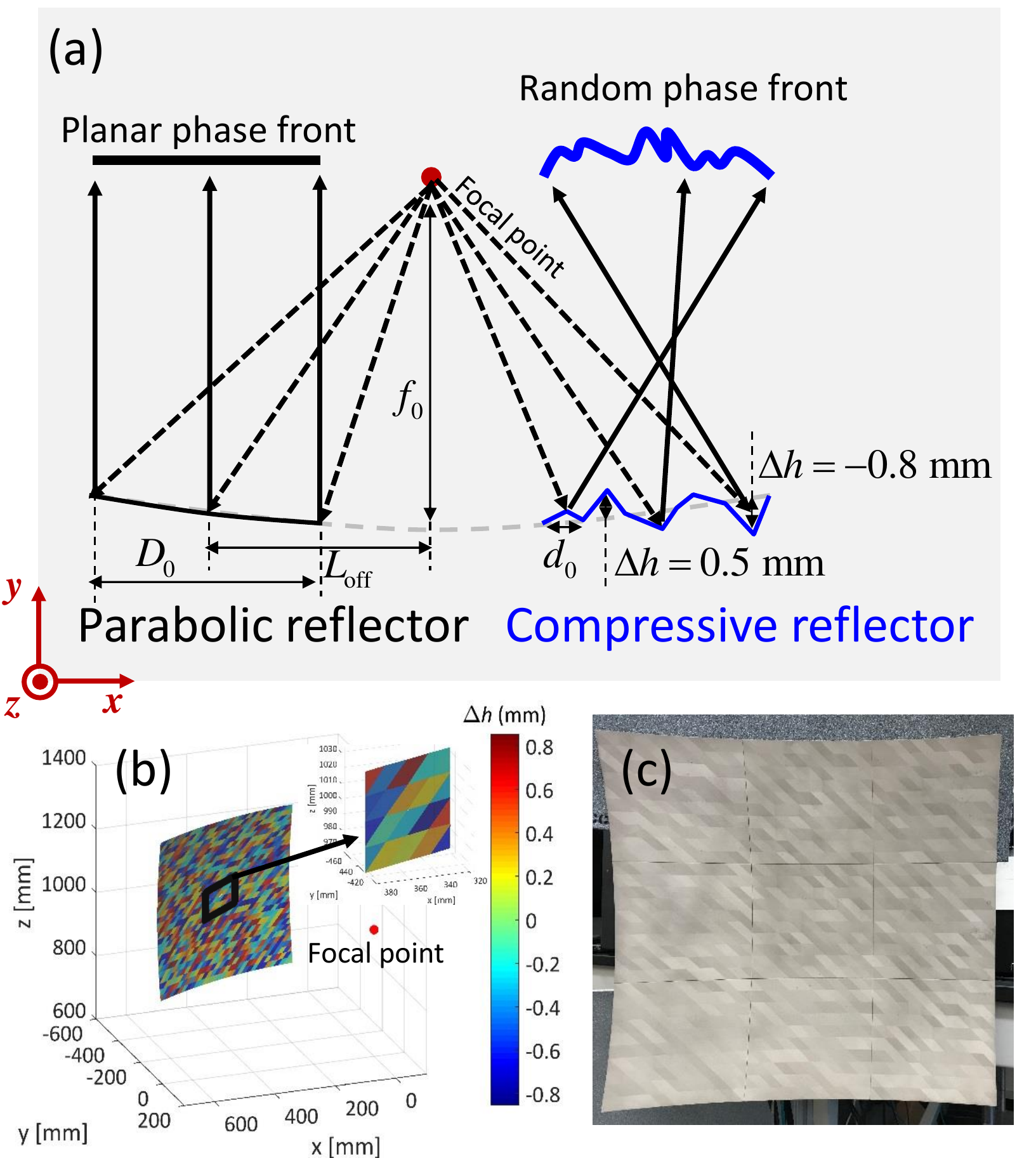}
\caption{(a) Off-axis placed parabolic (left) and compressive (right) reflector; (b) distortion hight $\Delta h$ between the CRA surface and TRA surface; and (c) fabricated CRA using 3D printing technique with its surface covered with metallic triangle units.}
\label{CRA_design}
\end{figure}
The surface of a TRA, which has its axis of symmetry parallel to the $\hat y $-axis, can be mathematically written as follows:
\begin{eqnarray}
\begin{aligned}
y = \frac{{{(x-L_\mathrm{off})^2} + {z^2}}}{{{4f_0}}} - {f_0}\\ \left\{ {x-L_\mathrm{off},z} \right\}\in[-D_0 /2, D_0 /2]
\end{aligned}
\label{eq_reflector}
\end{eqnarray}
where $D_0$ is the aperture size, ${f_0}$ is the focal length, and $F=(0,0,0)$ is the focal point that is co-located with the phase center of the MIMO array. In order to minimize the blockage of the reflected field from the TRA and to enhance the radiation efficiency, its surface is designed with an offset length of $L_\mathrm{off}$, as shown in the left part of Fig. \ref{CRA_design}(a). The coating of the TRA surface with metallic applique scatterers results in a CRA that performs the spatial coding of the fields reflected from its surface. The metallic applique scattererers can be described by a pseudo-random perturbation function in $\Delta h(x,z)$ in the $\hat y $-axis. A geometrical model of the CRA is plotted in the right part of Fig. \ref{CRA_design}(a), where the support of the $\Delta h(x,z)$ function has been discretized into a tessellated mesh of $N_{\Omega}$ triangular faces. The averaged side length of the triangular facets is $d_0$; and the distortion for each vertex  is drawn from an uniform random distribution $U(-\Delta h_m,+\Delta h_m)$, being $\Delta h_m$ the maximum allowed distortion. All the parameters of the CRA designed in this work are shown in Table I, where $\lambda_0$ is the wavelength corresponding to the center frequency $f_c$ of the radar that is operating with a bandwidth $B$. The exact distortion heights  of all the applique scatterers covering the CRA surface is plotted in Fig. \ref{CRA_design}(b)---see \cite{7289386,7928846} for a more in-depth description of CRA's parameters.
\begin{table}[!t]
\renewcommand{\arraystretch}{1.5}
\caption{Parameters of the CRA}
\centering

    \begin{tabular}{|c|c||c|c|}
\hline
Name & Value    & Name    & Value \\
\hline
$D_0$& $500$ mm & $\Delta h_m$ & $0.8$ mm \\
\hline
$L_\mathrm{off}$& $350$ mm & $\lambda_0$ & $4.1$ mm \\
\hline
 $f_0$ & $500$ mm & $B$ &  $5$ GHz\\
 \hline
$d_0$ & $16.4$ mm & $f_c$ & $73.5$ GHz\\
\hline
    \end{tabular}
\end{table}

The fabrication of the CRA is done using additive manufacturing techniques. Specifically, an Object Eden 260VS 3D printer is utilized with VeroWhitePlus as the filament material. Due to limitations on printing size, the CRA is divided into $9$ parts that are independently printed and assembled together afterwards. The surface of each part is metalized using an acrylic-based silver conductive coating spray. Figure \ref{CRA_design}(c) shows the manufactured CRA that is used to perform the wave-field compression and imaging discussed in the next section.

\begin{figure}[htp]
	\centering
	\includegraphics[width=\linewidth]{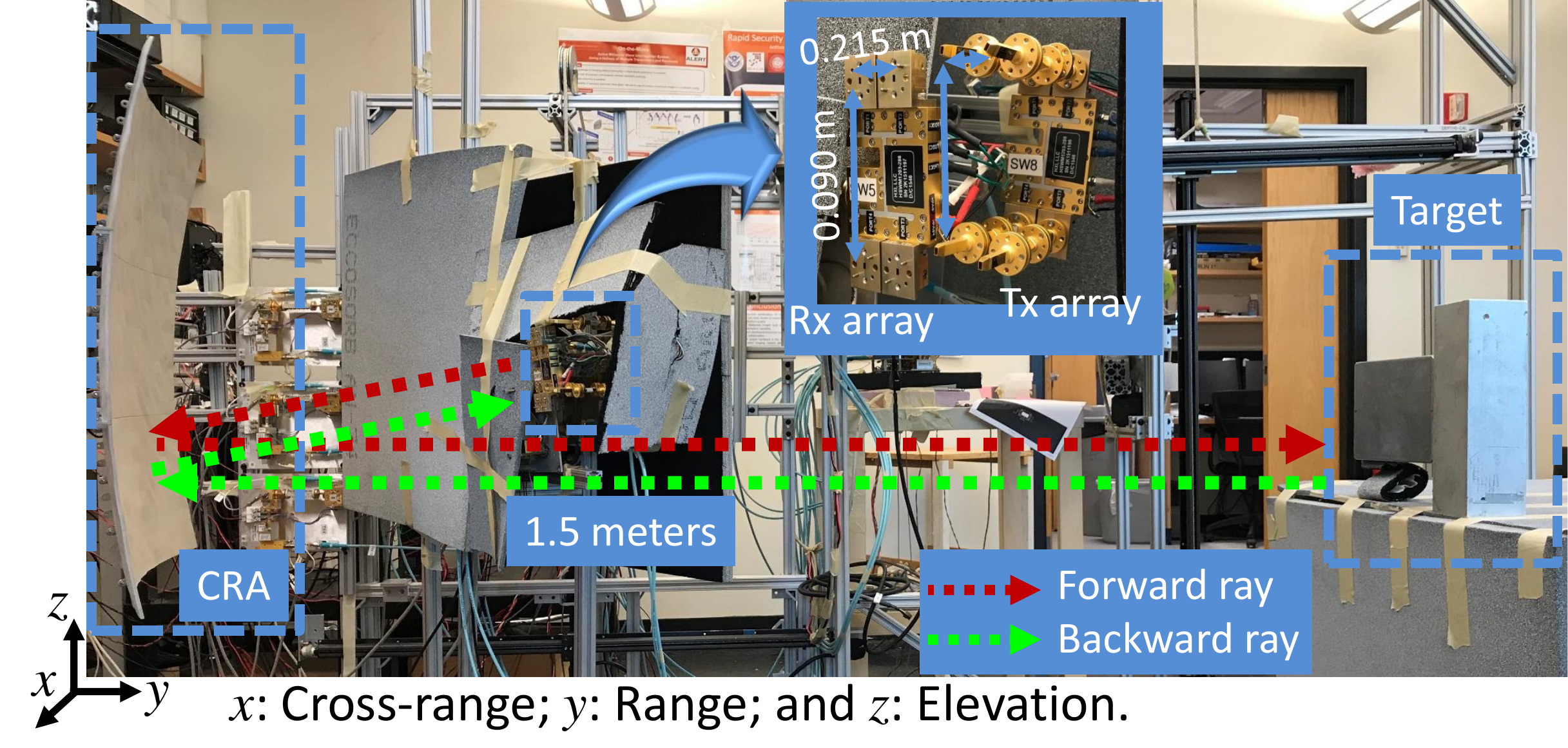}
	\caption{Experimental setup of the CRA imaging. The forward and backward paths are illustrated in the red and green dash-line, respectively. A T-shaped metallic target is located $1.5$ m away from the center of the CRA. The arrangement and dimensions of the Tx/Rx arrays are shown in the subplot.}
	\label{EXP_setup}
\end{figure}

The MIMO-array of the CRA uses a frequency modulated continuous wave (FMCW) radar front-end (RFE), developed by HXI\cite{HXI2012Rosa}. The operating frequency range of the RFE is from $70$ to $77$ GHz, however, only the frequencies from $71$ to $76$ GHz are used. The radar configuration employs one static Tx and one static Rx module. Each module has a single-pole four-throw (SP4T) switch, as shown in the subplot of Fig. \ref{EXP_setup}. The SP4T switches are driven by a FPGA switching system based on an Altera Cyclone V DE1-SoC board.

\section{Calibration of the CRA-based Sensing Matrix}

Ensuring phase coherence among all transmitting and receiving elements of the MIMO array poses an important challenge for this type of high-frequency imaging systems. This is due to the fact that small uncertainties on the relative position of each module--just a fraction of a millimeter--or on the transfer function will hinder the system's ability to perform a successful imaging. In this work, the phase coherence is ensured through a calibration step, in which the field distribution of each array element is measured on an aperture in front of the radar, and this data is later used to build the sensing matrix for one or more RoIs. The field distribution of each array element is collected in a two-step procedure. First, several FMCW chirps are sequentially transmitted from each one of the static MIMO transmitters, and they are measured on the 2D aperture with a moving receiver. Second, several FMCW chirps are transmitted from a moving transmitter on the same 2D aperture, and they are measured by each static receiver of the MIMO array.

The recorded fields corresponding to the static transmitting (Tx) and receiving (Rx) modules are denoted as ${\bf{E}}_{{\rm{tx}}}^{{\rm{apert}}}$ and ${\bf{E}}_{{\rm{rx}}}^{{\rm{apert}}}$, respectively; and they are used to construct a calibrated sensing matrix. First, the equivalent magnetic currents $\mathbf{M}_{\mathrm{tx}}^{\mathrm{aper}}$ and $\mathbf{M}_{\mathrm{rx}}^{\mathrm{aper}}$, corresponding to the transmitters and receivers, respectively, are computed using the equivalence theorem as follows \cite{Balanis2012Advanced}:
\begin{eqnarray}
\begin{aligned}
{\bf{M}}_{{\rm{tx/rx}}}^{{\rm{apert}}} &=  - 2{{\bf{n}}_0} \times {\bf{E}}_{{\rm{tx/rx}}}^{{\rm{apert}}},
\end{aligned}
\label{eq_current}
\end{eqnarray}
where ${{\bf{n}}_0} = [0,1,0]$ is the normal vector on the 2D aperture. Then, the incident fields in the RoI produced by these currents are computed using the exact near-field radiation equations\cite{Balanis2012Advanced} as follows:
\begin{eqnarray}
\begin{aligned}
{\bf{E}}_{{\rm{tx/rx}}}^{{\rm{RoI}}}\left( {\bf{r}} \right) &=  - \frac{1}{{4\pi }}\int_{{S_{{\rm{apert}}}}} {{G_0}\left( {{\bf{M}}_{{\rm{tx/rx}}}^{{\rm{apert}}} \times {\bf{R}}} \right){e^{ - j{k}R}}} ds,
\end{aligned}
\label{eq_near_gen}
\end{eqnarray}
where ${G_0} = \frac{{1 + j{k}R}}{{{R^3}}}$; ${\bf{R}} = {\bf{r}} - {\bf{r'}}$ with ${\bf{r}}$ being the point in the RoI and ${\bf{r'}}$ being the point on the 2D aperture; $R = \left| {\bf{R}} \right|$; $k = \frac{2\pi}{\lambda}$ is the wave number; $S_{\mathrm{apert}}$ is the surface of the 2D aperture; and $\mathbf{E}_{\mathrm{tx}}^{\mathrm{RoI}}$ and $\mathbf{E}_{\mathrm{rx}}^{\mathrm{RoI}}$ are the electric fields, generated by $\mathbf{M}_{\mathrm{tx}}^{\mathrm{apert}}$ and $\mathbf{M}_{\mathrm{rx}}^{\mathrm{apert}}$, respectively, in each voxel of the RoI. Finally, the first order Born approximation\cite{poli2012microwave} can be used to compute the calibrated sensing matrix for each transmitting and receiving port $\mathbf{H}_{T,R}$:
\begin{equation}
\mathbf{H}_{T,R} = \mathbf{E}_{\mathrm{tx}}^{\mathrm{RoI}} \cdot \mathbf{E}_{\mathrm{rx}}^{\mathrm{RoI}}
\end{equation}
Finally, the concatenation of all sensing matrices for each pair of transmitting and receiving ports and for all sampled frequencies give rise to the full-system sensing matrix $\mathbf{H}$ that will be used for the imaging.

\section{Imaging Algorithm}
The imaging algorithm is used to solve the Born-linearized inverse problem: $\mathbf{H}{\bf{u}}+{\bf{n}}={\bf{g}}$, where $\mathbf{u}$ is the unknown reflectivity vector of the target in the RoI; ${\bf{g}}$ is the measured field scattered by the target; and ${\bf{n}}$ is the noise vector. This problem can be iteratively solved with typical norm-$1$ regularized algorithms; and, in this work, the distributed alternating direction method of multipliers (ADMM)\cite{Juan2017Norm1ADMM} is applied as follows:
\begin{equation}
\begin{array}{l}
{\rm{minimize}} \; \frac{1}{2}\sum\limits_{i = 1}^N {\left\| {{{\bf{H}}_i}{{\bf{u}}_i} - {{\bf{g}}_i}} \right\|_2^2}  + \lambda_r {\left\| {\bf{v}} \right\|_1}\\
{\rm{s}}.{\rm{t}}. \;\;\;\;\;\;\;\;\;\; {\bf{u}}_i = {\bf{v}},\forall i = 1,...,N,
\end{array}
\end{equation}
where the original sensing matrix $\mathbf{H}$ and the measurement vector $\mathbf{g}$ are divided (by rows) into $N$ submatrices $\mathbf{H}_i$ and $N$ subvectors $\mathbf{g}_i$, respectively, $i \in [1, N]$; $\mathbf{u}_i$ is the unknown reflectivity vector for each pair of $\mathbf{H}_i$ and $\mathbf{g}_i$; $\lambda_r$ is the norm-$1$ weight factor; and ${\bf{v}}$ is a \textit{consensus} variable that enforces the agreement among all $\mathbf{u}_i$.

In order to mitigate the spatial noise, a 2D cross-range averaging was applied to the reflectivity function $\mathbf{u}$. The averaging process can be written as follows
\begin{equation}
\mathbf{u}_{a}({x_0},{y_0},z_0) = \sum\limits_{n_x =  - \frac{N_a}{2}}^{\frac{N_a}{2}} {\sum\limits_{n_z =  - \frac{N_a}{2}}^\frac{N_a}{2} {\frac{\mathbf{u}({x_0} + n_x,{y_0}, z_0 + n_z)}{{{(N_a+1)^2}}}} },
\label{eq_averaging_processing}
\end{equation}
where $\mathbf{u}({x_0},{y_0},z_0)$ is the reflectivity of the $({x_0},{z_0})$-th pixel in the $y_0$-th 2D plane, and the length of the 2D averaging process is $N_a+1$ in each dimension, $N_a$ being an even number.

\section{Experimental Results}

\begin{figure}[t]
	\centering
	\includegraphics[width=\linewidth]{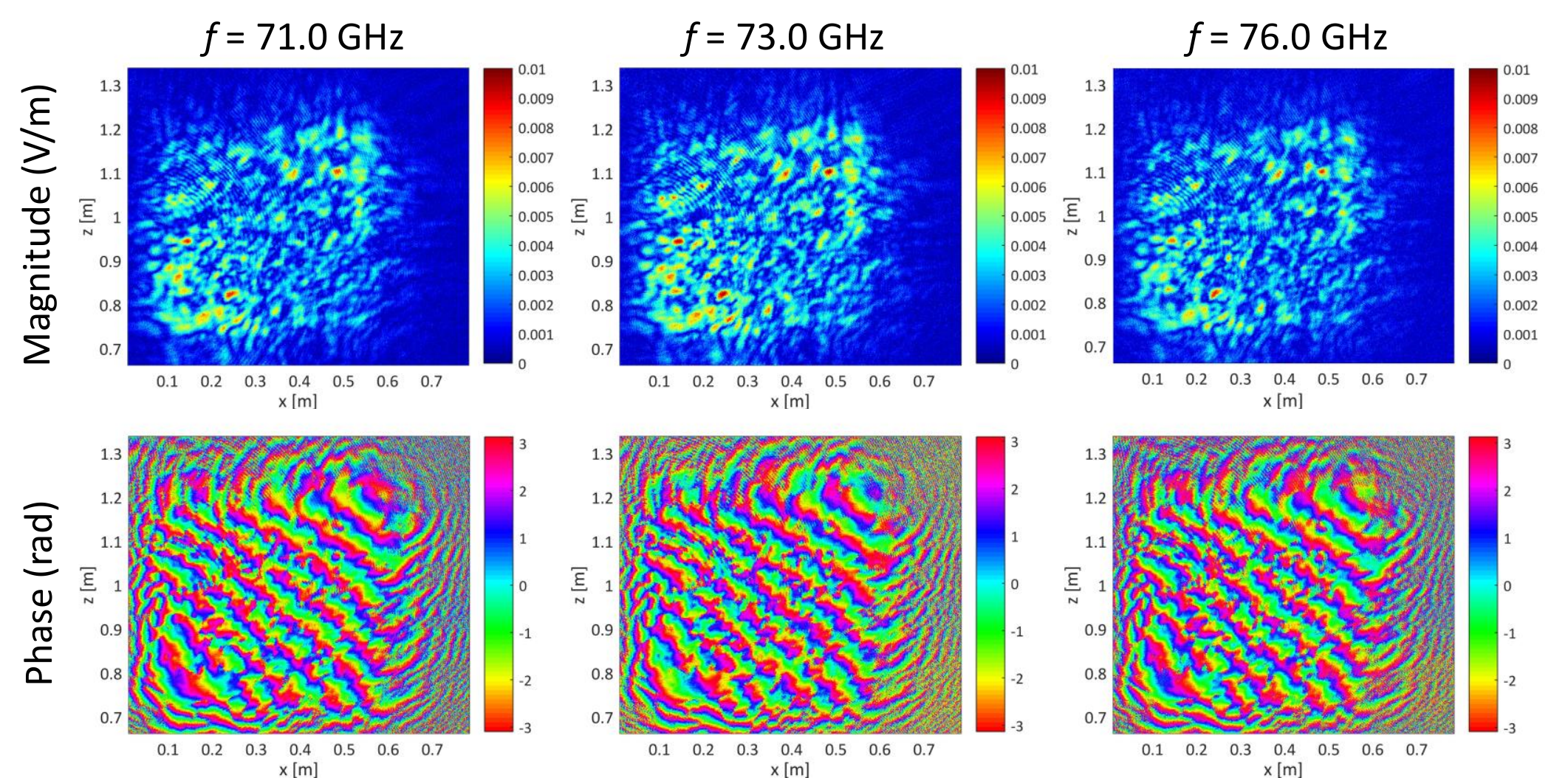}
	\caption{Measured near-field radiation patterns including the magnitude (top) and phase (bottom) distributions.}
	\label{CRA_near_field}
\end{figure}

The spatial codes created by the fabricated CRA are measured in the near field for the calibration algorithm all throughout the operating bandwidth. Two WR-12 tapered waveguides are used with the moving transmitter and receiver, which are mounted on the scanning platform at $900$ mm away from the CRA's center. The fields are collected in an aperture of $880$ mm and $640$ mm in the $x$-axis (cross-range) and the $z$-axis (elevation), respectively, and sampled every half a wavelength at $77$ GHz. The magnitude and phase of the measured fields for one of the MIMO transmitters are plotted for different frequencies in Fig. \ref{CRA_near_field}, showing the expected spatially-coded patterns.

 The experimental setting shown in Fig. \ref{EXP_setup} is used to image a rotated T-shaped metallic target. In the forward path (illustrated with red dashed-line), the four transmitting ports of the switch are sequentially used to illuminate the CRA. The incident wavefront is pseudo-randomly coded and scattered towards the RoI, where the T-shaped target is placed at its center. In the backward path (illustrated with the green dashed-line), the scattered field produced by the target is coded again by the CRA, and the field is scatted towards and measured at all four receiving ports. A total of $30$ evenly-spaced frequencies are used in the $5$ GHz bandwidth; resulting in a total of $480$ measurements ($30$ frequencies $\times$ $4$ Txs $\times$ $4$ Rxs). The dimensions of the RoI are $600$ mm, $420$ mm, and $600$ mm in cross-range, range, and elevation, respectively. The voxel discretization size is selected to be $6$ mm, $30$ mm, and $6$ mm in $x-$, $y-$, and $z-$axis, respectively, which is limited by the upper bound range and cross-range resolution of wideband radar aperture systems \cite{Sheen2001Three}:
\begin{eqnarray}
\begin{aligned}
{\sigma _{x,z}} &= \frac{{{\lambda _0}R}}{{2{D_0}}},\\
{\sigma _y} &= \frac{c}{{2B}},
\end{aligned}
\label{eq_reflector}
\end{eqnarray}
where $R = 1.5$ meters is the range of the target, $D_0 = 500$ mm is the aperture size of the reflector, $\lambda_0$ is the free-space wavelength corresponding to the center frequency of $73.5$ GHz, $c$ is the speed of light, and $B$ = 5 GHz is the operating bandwidth of the radar system.

\begin{figure}[t]
\centering
\includegraphics[width=0.5\linewidth]{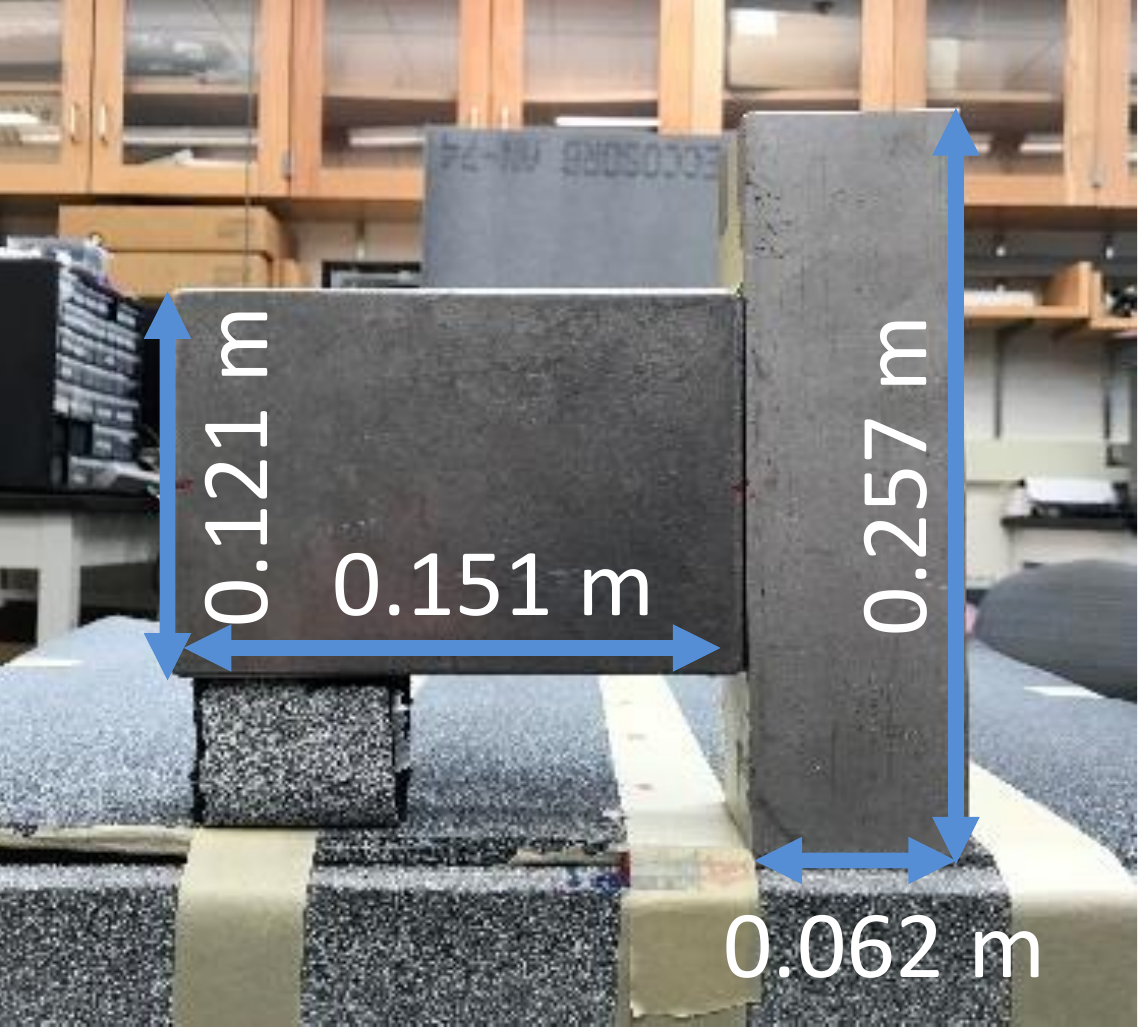}
\caption{T-shaped metallic target under detection.}
\label{EXP_target}
\end{figure}
\begin{figure}[t]
\centering
\includegraphics[width=\linewidth]{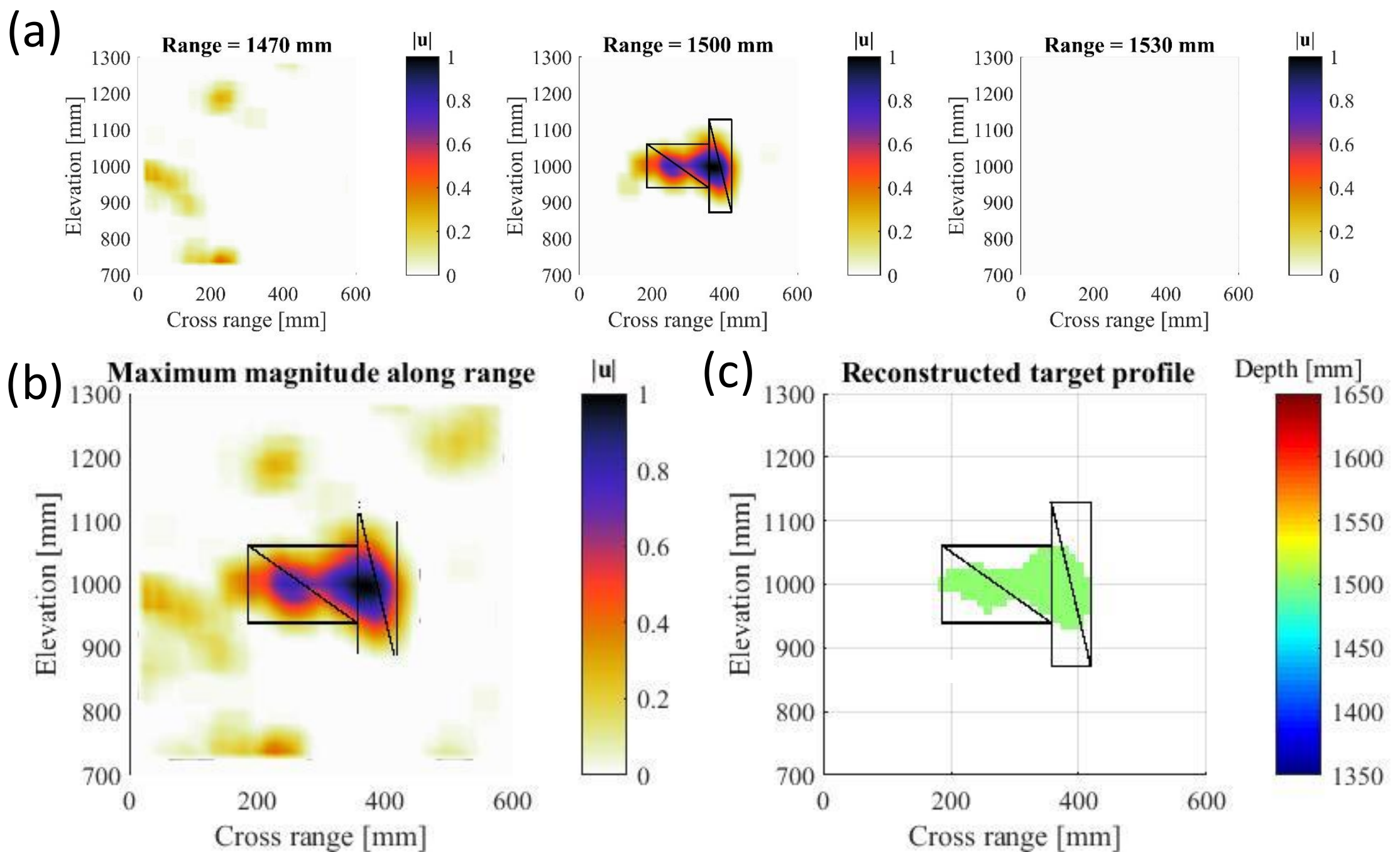}
\caption{(a) Normalized magnitude of the reconstructed reflectivity, $|\mathbf{u}|$, on the 2D planes at different ranges; (b) maximum $|\mathbf{u}|$ along the range ($y$-axis); and (c) reconstructed target profile with a display threshold of $|\mathbf{u}|\geq0.35$. The 2D averaging process with a length of $N = 5$ pixels is applied.}
\label{EXP_result}
\end{figure}
The dimensions of the rotated T-shaped metallic target under test are shown in Fig. \ref{EXP_target}. The norm-$1$ regularized ADMM algorithm described above is executed--using a weight factor of $\lambda_r = 20$ and dividing the sensing matrix into $N=40$ blocks. A $5$-pixel 2D averaging processing ($N_a=4$) is applied afterwards. The 3D reconstruction of the target profile is plotted in Fig. \ref{EXP_result}. Three out of fifteen imaged frames of the reconstructed reflectivity magnitude $|\mathbf{u}|$ are shown in Fig. \ref{EXP_result}(a); the maximum magnitude of $|\mathbf{u}|$ along the range ($y$-axis) is plotted in Fig. \ref{EXP_result}(b); and the reconstructed target profile is shown in Fig. \ref{EXP_result}(c) where a display threshold of $|\mathbf{u}|\geq0.35$ has been applied. These figures show that the CRA-based system is capable of imaging a target in the RoI. Although several artifacts appear at ranges closer to the radar system, the target profile is reconstructed and shows the expected T-shaped profile given a suitable amplitude threshold. The difference between the reconstructed image and the real target can be mainly attributed to the noisy measurements of the sensing matrix $\mathbf{H}$ and the vector $\mathbf{g}$, as well as the non-coaxial alignment between the target and the CRA.

\section{Conclusion}
This paper presents the first experimental results of our 3D mm-wave Compressive-Reflector-Antenna-based imaging system. The proposed cost-effective imaging system is built using a 3D-printed CRA fed by a MIMO array that uses $4$ transmitting and $4$ receiving ports. The CRA is used to introduce spatial and spectral coding of the wavefields incident to and scattered from the target, thus enhancing the diversity of successive measurements. The experimental results showed that the proposed system is capable of reconstructing the shape and reflectivity function of the target under test. This paper paves the way towards developing a new cost-efficient mm-wave imaging system that uses multiple CRAs to image on-the-move human-size targets in security applications.

\section*{Acknowledgment}
This work has been partially funded by the NSF CAREER program (Award No. 1653671) and the U.S. Department of Homeland Security (Award No. 2013-ST-061-ED0001).

\bibliographystyle{IEEEtran}
\bibliography{myreference}

\end{document}